\newcommand{\be}{\begin{equation}}
\newcommand{\ee}{\end{equation}}
\newcommand{\bea}{\begin{eqnarray}}
\newcommand{\eea}{\end{eqnarray}}
\newcommand{\nn}{\nonumber \\}
\newcommand{\p}[1]{(\ref{#1})}
\newcommand{\lb}[1]{\label{#1}}
\newcommand\s{\scriptscriptstyle}

\newcommand\q{\quad}
\newcommand\qq{\quad\quad}

\newcommand\cA{{\cal A}}

\newcommand\cN{{\cal N}}

\newcommand\cW{{\cal W}}

\newcommand\olQ{\overleftarrow{Q}}
\newcommand\orQ{\overrightarrow{Q}}

\newcommand\Tr{\mbox{Tr}\,}

\newcommand\tpa{\theta^{+\alpha}}

\newcommand\tma{\theta^{-\alpha}}

\newcommand\tka{\theta^{\alpha}_k}

\newcommand\btka{\bar\theta^{\da k}}

\newcommand\btja{\bar\theta^{\da j}}

\newcommand\btpa{\bar{\theta}^{+\dot{\alpha}}}

\newcommand\btm{\bar{\theta}^-}

\newcommand\btma{\bar{\theta}^{-\dot{\alpha}}}

\newcommand\tp{\theta^+}
\newcommand\tm{\theta^-}
\newcommand\btp{\bar\theta^+}

\def\a{\alpha}
\def\da{{\dot\alpha}}
\def\b{\beta}
\def\db{{\dot\beta}}
\def\g{\gamma}

\def\d{\delta}

\def\eps{\epsilon}

\def\ve{\varepsilon}

\def\bph{{\bar\phi}}
\def\vp{\varphi}

\def\j{\psi} \def\bj{{\bar\psi}}
  
\def\l{\lambda}
 \def\th{\theta}  \def\bt{\bar\theta}

\def\si{\sigma}

\def\bs{\bar\sigma}

\def\J{\Psi}
\def\bJ{\bar\Psi}
\def\L{\Lambda}

\def\pa{\partial}
\def\na{\nabla}

\newcommand\ada{{\alpha\dot{\alpha}}}

\newcommand\bdb{{\beta\dot{\beta}}}

\newcommand\ab{{\alpha\beta}}

\newcommand\pada{\partial_{\alpha\dot{\alpha}}}

\newcommand\A{{\s A}}

\newcommand\C{{\s C}}

\newcommand\R{{\s R}}

\newcommand\sL{{\s L}}

\newcommand\W{{\s W}}
\newcommand\Z{{\s Z}}

\newcommand{\0}{{\s 0}}

\newcommand{\pp}{{\s ++}}
\newcommand{\m}{{\s --}}

\newcommand{\Dp}{D^{\pp}}
\newcommand{\Dm}{D^{\m}}

\newcommand{\dpp}{\partial^{\pp}}

\newcommand{\Vp}{V^\pp}
\newcommand{\Vm}{V^\m}

\newcommand{\Dpa}{D^+_\alpha}
\newcommand{\Dma}{D^-_\alpha}

\newcommand{\bDpa}{\bar{D}^+_{\dot{\alpha}}}

\newcommand{\bDma}{\bar{D}^-_{\dot{\alpha}}}

\def\sfrac#1#2{{\textstyle\frac{#1}{#2}}}

\def\sha{\sfrac12}

\def\e{\mbox{e}}
\def\ii{\mbox{i}}
\def\diff{\mbox{d}}

\documentclass[12pt]{article}
\usepackage{amsmath,amssymb}
\renewcommand{\theequation}{\thesection.\arabic{equation}}
\csname @addtoreset\endcsname{equation}{section}

\def\theequation{\arabic{section}.\arabic{equation}}
\begin{document}
\begin{titlepage}

\begin{flushright}
CERN-PH-TH/2004-032\\
ITP-UH-10/04\\
LAPTH-1041/04\\
hep-th/0405049
\end{flushright}

\begin{center}
{\bf NON-ANTICOMMUTATIVE CHIRAL SINGLET DEFORMATION \\[8pt]
OF N=(1,1) GAUGE THEORY}
\vspace{1cm}

{\bf S. Ferrara$\,{}^a$, E. Ivanov$\,{}^b$, O. Lechtenfeld$\,{}^c$, 
E. Sokatchev$\,{}^d$, B. Zupnik$\,{}^b$}
\vspace{0.7cm}

${}^a${\it CERN, Theory Division, CH 1211, Geneva 23, Switzerland;\\
INFN, Laboratori Nazionali di Frascati, I-00044 Frascati Italy}\\
{\tt sergio.ferrara@cern.ch }\\[8pt]
${}^b${\it Bogoliubov Laboratory of Theoretical Physics,
JINR, \\
141980, Dubna, Moscow Region, Russia}\\
{\tt eivanov, zupnik@thsun1.jinr.ru}\\[8pt]
${}^c${\it Institut f\"ur Theoretische Physik, Universit\"at Hannover, \\
30167 Hannover, Germany}\\
{\tt lechtenf@itp.uni-hannover.de}\\[8pt]
${}^d${\it Laboratoire d'Annecy-le-Vieux de Physique 
The\'orique\footnote{UMR 5108 
assosi\'ee
\`a l'Universit\'e de Savoie} LAPTH, \\
Chemin de Bellevue
- BP 110 - F-7491 \\
d'Annecy-le-Vieux Cedex, France}\\
{\tt emeri.sokatchev@cern.ch}
\end{center}
\vspace{1cm}

\begin{abstract}
\noindent We study the SO(4)$\times$SU(2) invariant Q-deformation of 
Euclidean $N{=}(1,1)$ gauge theories in the harmonic superspace formulation.
This deformation preserves chirality and Grassmann harmonic analyticity
but breaks $N{=}(1,1)$ to $N{=}(1,0)$ supersymmetry.
The action of the deformed gauge theory is an integral over the chiral
superspace, and only the purely chiral part of the covariant superfield 
strength contributes to it.
We give the component form of the $N{=}(1,0)$ supersymmetric action
for the gauge groups U(1) and U($n{>}1$). 
In the U(1) and U(2) cases, we find the explicit nonlinear field redefinition 
(Seiberg-Witten map) relating the deformed $N{=}(1,1)$ gauge multiplet 
to the undeformed one. This map exists  in the general U($n$) case as well,
and we use this fact to argue that the deformed U($n$) gauge theory can be
nonlinearly reduced to a theory with the gauge group SU($n$).

\end{abstract}
\end{titlepage}

\setcounter{page}{1}
\section{Introduction}

Currently, deformations of supersymmetric field theories (for a review 
see \cite{Douglas:2001ba}) are under intensive study 
in the framework of deformed superspace 
(see, e.g., \cite{FL}-\cite{FLM}). The anticommuting deformations
\cite{Brink:nb}-\cite{deBoer:2003dn} are
related to superstring theories in particular backgrounds 
with non-vanishing Ramond-Ramond
field strengths \cite{OoVa,Se}.

An important class of nilpotent Q-deformations of the Euclidean 
superfield $N{=}(\sfrac12,\sfrac12)$ 
theories was considered in \cite{Se}. These deformations are 
generated by the  
Poisson bi-differential operator
$P$ constructed in terms of half of the supersymmetry generators 
($N{=}(\sfrac12,0)$ in our
notation). This operator is nilpotent, $P^3=0$, and it breaks the 
$N{=}(0,\sfrac12)$ supersymmetry, 
still preserving chirality. An alternative type of nilpotent deformations, the 
D-deformations, uses Poisson operators which are bilinear 
in the spinor covariant 
derivatives and so preserve the full supersymmetry \cite{FL,FLM}.

The $N{=}(1,1)$ generalizations of both Q- and D-nilpotent deformations
were analyzed in the framework of the harmonic superspace approach 
in \cite{ILZ,FS}.\footnote{ Note that the bosonic deformations of 
$N{=}2$ harmonic superspace have been considered earlier in \cite{BS}.} 
In particular, in \cite{FS}
the simplest D-deformation of $N{=}(1,1)$ gauge theory was studied, 
the singlet one which preserves
both supersymmetry and SU(2)$_\sL\times$SU(2)$_\R\times$SU(2) symmetry. 
It also preserves Grassmann
harmonic analyticity and chirality (but not anti-chirality).

The Q-deformations of the chiral or harmonic $N{=}(1,1)$ superspaces 
generically break $N{=}(0,1)$ supersymmetry and the SU(2)$_\sL\times$
SU(2)$_\R\times$O(1,1) symmetry. However, there exists 
an interesting Q-singlet (QS) deformation
with the following SU(2)$_\sL\times$SU(2)$_\R\times$SU(2) invariant 
Poisson structure,
\be
P_s = -I \varepsilon^{\a\b}\varepsilon_{ij} 
\overleftarrow{Q}_{\a}^i \overrightarrow{Q}_{\b}^j\,. \lb{Psing}
\ee
It corresponds to the Moyal-Weyl star product
\be
  A\star B = A\,\exp\left(-I \varepsilon^{\a\b}\varepsilon_{ij} \olQ_\a^i
  \orQ_\b^j \right) B\,.
\label{1}
\ee
Here $Q^k_\a$ are the $N{=}(1,0)$ supersymmetry generators,  
$I$ is the constant deformation parameter,
$A$ and $B$ are some arbitrary superfields. While breaking half 
of $N{=}(1,1)$ supersymmetry,
this deformation preserves both types of chirality, 
as well as Grassmann harmonic analyticity \cite{ILZ,FS}.

In this paper we study the properties of the QS-deformed U($n$) gauge theories.
In particular, we find the explicit component field actions for these theories.
They can be read off directly from the deformed harmonic superfield 
actions \cite{ILZ,FS}, using the WZ gauge
for the basic $N{=}(1,1)$ gauge potential in close analogy 
with the gauge theories in
the undeformed harmonic superspace \cite{GIKOS,GIOS}. 
The basic conventions
of the Euclidean harmonic $N{=}(1,1)$ superspace are reviewed in the Appendix.

Section 2 is devoted to the analysis of the U(1) gauge and supersymmetry 
transformations
of the component fields using the WZ gauge for the Grassmann analytic 
superfield potential. The chiral part $\cA$ of the covariantly chiral 
deformed superfield strength is 
constructed in Section 3. It is shown that the gauge invariant action 
can be expressed 
as the chiral superspace integral of the superfield $ \cA^2$. 
We also derive the equivalence
transformation relating the component fields of the undeformed 
and deformed U(1) gauge 
theories (the analog of the Seiberg-Witten transform \cite{SW}). 
The resulting deformed 
component action is surprisingly simple: 
\be
S = \int\!\diff^4x\ (1 +4I\bar\phi)^2\, L_0\, ,\lb{ActionC'}
\ee
where $L_0$ is the standard undeformed Lagrangian of the $N{=}(1,1)$ 
supersymmetric U(1) 
gauge theory and $\bar\phi$ is one of the two scalar fields 
of the  $N{=}(1,1)$ gauge 
multiplet. Thus, the only effect of the nilpotent QS-deformation 
in the Abelian case is 
the appearance of the scalar factor in (\ref{ActionC'}) 
which breaks half of the original 
supersymmetry. 

The QS-deformation of the non-Abelian U($n$) theory is considered in 
Section 4.  The deformed component action consists of two parts. 
The first is similar to (\ref{ActionC'}), the 
fields in the standard Lagrangian being rescaled by a matrix scalar factor. 
The second part contains some higher-derivative interaction terms which, 
however, might be removable by suitable field redefinitions.
We demonstrate the existence
of a Seiberg-Witten-type map to the undeformed gauge multiplet in the
non-Abelian case as well and give it explicitly for the gauge group 
U(2).\footnote{
Nilpotent Q-deformations of the on-shell superfield constraints
of the Euclidean $N{=}(2,2)$ gauge theory were studied in~\cite{SaWo}.}
Due to this map, the deformed U($n$) gauge theory can be nonlinearly reduced
to the deformed SU($n$) one.

Finally, Section 5 relates the singlet deformation to a particular type 
of IIB string background of the Ramond-Ramond axion field strength.

Note that some lowest-order deformation terms in the Q-deformed U(1) 
component action were found in the recent paper \cite{AIO}, 
also making use of the harmonic superspace approach.\footnote{
When preparing our paper for submission to the e-archive, we became aware of 
the new article \cite{AIO2} where the component action and the supersymmetry 
transformations of the U(1) QS-deformed theory are also obtained.}
Preliminary results of this work have been reported at the 
International workshops ``Supersymmetry and string theory'' 
(Padova University, October 2-4, 2003) and  ``Classical and quantum 
integrable systems''  (Dubna, January 26-29, 2004).

\section{Generalities of QS-deformed N=(1,1) gauge theory} 

An overview of the Euclidean $N{=}(1,1)$ chiral and harmonic superspaces 
can be found in \cite{ILZ,FS}. For the convenience of the reader, however,
we have collected all relevant notation and definitions in the Appendix.

We shall study the SO(4)$\times$SU(2) invariant
deformation of the U(1) supersymmetric gauge theory which corresponds 
to the following
Poisson bracket of two arbitrary superfields $A$ and $B$:
\be
A P_s B=- I (-1)^{p(A)}Q^k_\a A \,Q^\a_k B\lb{PLbra}
\ee
where $P_s$ is the Q-singlet bi-differential operator, $I$ is 
the deformation constant,
$p(A)$ is the $Z_2$-grading,
and $Q^k_\a$ are the generators of $N{=}(1,0)$ supersymmetry 
given in \p{QLgen}. The operator $P_s$ is
nilpotent, $P^5_s=0$, it breaks down the $N{=}(0,1)$ supersymmetry 
and the O(1,1) automorphism group.
We shall use the nilpotency
of $P_s$ in  the basic formula for the non-commutative
product of superfields
\be
A\star B=:A\e^{P_s}B=A B+A P_sB+\sfrac12 AP^2_s B+\sfrac16 AP^3_sB 
+ \sfrac{1}{24}AP^4_sB\,.\lb{star}
\ee

This $\star$ product preserves both chirality ($\bar D_{\da l}\Phi=0$) 
and anti-chirality ($D^k_\a\Phi=0$):
\be
[D^k_\a,P_s]=[\bar D_{\da l},P_s]=0\,.
\ee
The analog of chirality in $N{=}(1,1)$ harmonic superspace is
Grassmann analyticity, $D^+_\a\Lambda=\bar D^+_\da\Lambda=0$. 
The differential condition for independence of the harmonic coordinates 
$u^\pm_k$ is  $D^{\pm\pm}A=0$. Both these conditions are preserved 
by the $\star$ product:
\be
[D^+_{\a},P_s]=[\bar D^+_{\da},P_s]=[D^{\pm\pm},P_s]=0\,.
\ee

The unconstrained Grassmann-analytic  potential of the 
Q-deformed U($n$) gauge theory
$V^{++}$ has the following gauge transformations,
\bea
&&\delta_\Lambda V^{++}=D^{++}\Lambda + [V^{++}, \Lambda]_\star\,,\nn
&&\widetilde{V^{++}}=-V^{++},\q\widetilde\Lambda=-\Lambda \lb{tilda1}
\eea
where $\Lambda$ is the infinitesimal analytic parameter  and
\bea
[V^{++}, \Lambda]_\star &=& [V^{++},\L]+I[Q^{+\a}V^{++}, Q^-_\a \Lambda]
-I[Q^{-\a} V^{++}, Q^{+}_{\a} \Lambda] \nn
&& -\, \sfrac14 I^2[(Q^+)^2V^{++},(Q^-)^2\L]-\sfrac14 I^2
[(Q^-)^2V^{++},(Q^+)^2\L] \nn
&& -\,I^2[Q^{-\b} Q^{+\a} V^{++},Q^{-}_\a Q^{+}_\b\L]\,,\lb{Abrack}
\eea
with $Q^\pm_\a=u^\pm_kQ^k_\a$ denoting the harmonic projections of 
the supersymmetry generators.
Note that the $I^3$ and $I^4$ terms of the deformed commutator vanish
on the full set of U($n$) algebra-valued analytic superfields. Also, 
the $I^0$ and $I^2$ terms vanish
for the deformed  U(1) superfields (i.e. in the Abelian case), 
so the U(1) gauge transformation
is
\bea
&&\delta_\Lambda V^{++}=D^{++}\Lambda+2IQ^{+\a}V^{++} Q^-_\a \Lambda
-2IQ^{-\a} V^{++} Q^{+}_{\a} \Lambda\,.
\lb{GaugetranV}
\eea

In what follows we shall use the chiral basis in $N{=}(1,1)$ superspace. 
It is best suited for a deformation which preserves chirality. In this basis
$$
Q^i_\alpha = \partial^i_\alpha =\partial/\partial \theta^\alpha_i\,, 
\quad Q^{\pm}_\alpha = \pm \partial_{\mp\a}\,,
\;\;\partial_{\pm \alpha} \equiv \partial /\partial \theta^{\pm\alpha}
$$
and
\be
P = -I \varepsilon^{\alpha\beta}\varepsilon_{kj}
\overleftarrow{\partial}^k_\alpha \overrightarrow{\partial}^j_\beta =
I\left(\overleftarrow{\partial}^\alpha_{+}\overrightarrow{\partial}_{-\alpha} -
\overleftarrow{\partial}^\alpha_{-}
\overrightarrow{\partial}_{+\alpha}\right)\,.
\ee
Then  the powers of the deformation operator appearing in \p{star}
can be written in a more explicit form as follows,
\bea
AP_sB&=&(-1)^{p(A)}I\,(\partial^\a_+ A \partial_{-\a} B - 
\partial_{-}^\a A \partial_{+\a}B)\,,\nn
\sfrac12 AP^2_sB&=&-\sfrac14 I^2\left[(\pa_+)^2A(\pa_-)^2B + (\pa_-)^2A
(\pa_+)^2B\right]
-I^2\pa_{+}^\b \pa^\a_-A\pa_{+\a} \pa_{-\b}B\,,\nn
\sfrac16 AP^3_sB&=&(-1)^{p(A)}\sfrac14 I^3\left[\pa^\a_-(\pa_+)^2A \pa_{+\a}
(\pa_-)^2B
- \pa^\a_{+}(\pa_-)^2A \pa_{-\a}(\pa_+)^2B\right], \nn
\sfrac{1}{24}AP^4_sB&=&
\sfrac{1}{16}I^4(\pa_+)^2(\pa_-)^2A (\pa_-)^2(\pa_+)^2 B\,. 
\lb{12}
\eea
Here $A$ and $B$ are arbitrary superfields and 
$(\partial_\pm)^2 = \partial^\a_{\pm}\partial_{\pm\a}$.

The basic object used for constructing invariant actions is the non-analytic
harmonic gauge potential $V^{--}$.
It is related to $V^{++}$ by the equation
\be
D^{++}V^{--} - D^{--}V^{++} + [V^{++}, V^{--}]_\star = 0
\,.\lb{hzc}
\ee
Its gauge transformation is
\be
\delta_\L V^{--} = D^{--}\Lambda + [V^{--}, \Lambda]_\star\,.\lb{GaugetranV2}
\ee
The harmonic zero-curvature equation for the non-analytic harmonic 
connection $V^{--}$
\p{hzc} has a non-polynomial superfield solution in the general gauge
\cite{GIOS,Z1}.
The corresponding superfield action is a non-polynomial 
(and non-local with respect to the
 harmonic variables) function of the basic superfield $V^{++}$. This 
manifestly $N{=}(1,0)$ 
covariant superfield formalism can be
used for supergraph calculations in the deformed theory. 
To get insights into the physical implications  of the 
singlet deformation, it is important to explore in detail  
the field-component structure of this superfield model.
We shall first study the QS-deformed U(1) gauge model and then turn to
the more complicated U($n$) case.

\section{QS-deformed U(1) gauge theory}
\subsection{Wess-Zumino gauge}
We start with the Abelian case, i.e. the QS-deformation 
of the U(1) gauge theory.

The WZ gauge for the U(1) gauge potential $V^{++}$ with 
Grassmann analytic coordinates
is given by
\bea
V^{++}_{\W\Z}(\zeta,u)&=& (\tp)^2\bph(x_\A)+(\btp)^2\phi(x_\A)+2(\tp\si_m\btp)
A_m(x_\A) + 4(\btp)^2\theta^+\Psi^-(x_\A)\nn
&&+ \,4(\tp)^2\btp \bar\Psi^-(x_\A)
+3(\tp)^2(\btp)^2{\cal D}^{--}(x_\A)\lb{WZanal}
\eea
where $\zeta=(x^m_\A, \tpa, \btpa)$ and
\be
\Psi^-_\a(x_\A) = \Psi^k_\a(x_\A)u^-_k\,, \quad \bar\Psi^{\da -}(x_\A) 
= \bJ^{\da k}(x_\A)u^-_k\,,
\quad {\cal D}^{--} = {\cal D}^{kl}(x_A)u^-_ku^-_l \lb{fields}
\ee
and the summation over repeated indices $\alpha$ and $\dot\alpha$ 
is defined in the standard way. The fields in \p{WZanal}, \p{fields}  
form the off-shell vector multiplet 
of $N{=}(1,1)$ supersymmetry.
To avoid a possible confusion, let us emphasize that the Euclidean 
fields $\phi(x_\A)$ and $\bar\phi(x_\A)$,
as well as $\Psi^k_\a(x_\A)$ and $\bJ^{\da k}(x_\A)$, 
are not mutually conjugated.

The analytic parameter of the residual gauge freedom has the following form:
\be
\L_r=\ii a(x_\A),\q Q^-_\a \L_r=-\pa_{+\a}\L_r=0\lb{resid}
\ee
where $a(x_\A)$ is some real function. The corresponding gauge 
transformation of
$V^{++}_{\W\Z}(\zeta)$ is given by \p{GaugetranV} with $\L = \L_r$.

It is convenient to use the chiral-analytic coordinates \p{Ccoor}.
With the help of the relations
$$
x^m_\A = x^m_\sL - 2\ii \theta^{-}\sigma^m\bar\theta^+
$$
and \p{leftanal} one can rewrite the gauge potential \p{WZanal} 
and the residual gauge freedom parameter \p{resid} as
\bea
&&V^{++}_{\W\Z}(z_\C, \bar\theta^+, u)=v^{++}(z_\C,u)+\btp_\da v^{+\da}(z_\C,u)
+(\btp)^2v(z_\C,u)\,,
\nn
&& \L_r=\ii a(x_\sL)+2\tm\si_m\btp\pa_ma(x_\sL)-
\ii(\tm)^2(\btp)^2\Box a(x_\sL)\,.
\lb{WZchir}
\eea
Here
$$
z_\C = (x_\sL^m,\tpa, \tma )
$$
and
\bea
v^{++} &=& (\theta^+)^2\,\bar\phi\,, \; v^{+\dot\alpha} 
\ =\ 2 (\theta^+ \sigma_m)^{\dot\alpha}\,A_m
+ 4(\theta^+)^2\,\bar\Psi^{-\dot\alpha} - 2\ii\,
(\theta^+)^2\,(\theta^- \sigma^m)^{\dot\alpha}\,\partial_m \bar\phi \,, \nn
v &=& \phi + 4\theta^+ \Psi^- + 3(\theta^+)^2\,{\cal D}^{--} 
-2\ii (\theta^+ \theta^-)\,\partial_m A_m
 -  \theta^- \sigma_{mn}\theta^+ \,F_{mn}\nn
&&-\, (\theta^+)^2 (\theta^-)^2\Box \bar\phi
+ 4 \ii\,(\theta^+)^2 \,\theta^-\si_{m}\partial_m \bar\Psi^-\,, \label{v++}\\
F_{mn} &=& \pa_m A_n-\pa_n A_m\,.\nonumber
\eea

The residual deformed U(1) gauge transformation is
\be
\delta_r V^{++}_{\W\Z}=D^{++}\L_r +
2I\pa^\a_{+} V^{++}_{\W\Z} \pa_{-\a} \L_r\,.\lb{00}
\ee
Then the gauge transformations of the component fields read
\bea
&&\delta_r\phi=-8I A_m\pa_m a~,\q \delta_r\bph=0\,,\q \delta_rA_m
=(1{+}4I\bph)\pa_m a\,,\nn
&&\delta_r\Psi^k_\a=-4I(\sigma_m \bar\Psi^{k})_\a \partial_m a \,,
\q \delta_r\bar\Psi^k_\da=0\,,\q
\delta_r {\cal D}^{kl}=0\,.\lb{defgauge}
\eea

It is straightforward to find an equivalence transformation 
which brings these deformed
gauge transformations to the standard form. This ``minimal'' 
Seiberg-Witten (SW) map takes the form
\bea
&& A_m = a_m(1 +4I\,\bar\phi)\,, \;\;
\phi  = \widetilde{\phi} - 4I a_m^2 (1 + 4I\,\bar\phi)\,, \;\;
\Psi^i_\alpha = \widetilde{\Psi}^i_\alpha - 4I\,
(\sigma_m\bar\psi^i)_\alpha a_m\,,\nn
&& F_{mn} = (1 +4I\,\bar\phi)f_{mn} +4I\,\left(\partial_m \bar\phi a_n -
\partial_n \bar\phi a_m\right)\, \lb{14}
\eea
where $f_{mn} = \pa_m a_n-\pa_n a_m$. It follows that 
\be
\delta_r a_m = \partial_m a\,, \;\; \delta_r \widetilde{\phi }
= \delta_r \widetilde{\Psi}^i_\alpha = 0\,.
\lb{15'}
\ee
Note that the gauge transformations of the redefined fields 
include the same gauge parameter as in \p{defgauge}, and this 
property is featured by the non-Abelian case as well. This is 
a difference from the standard bosonic SW map \cite{SW} which 
implies also a field-dependent redefinition of the gauge 
parameters.   

In fact all fields can be further redefined in such a way
that they will have the standard transformation properties 
under both the gauge and
$N{=}(1,0)$ supersymmetry transformations. We will postpone 
the discussion of this issue to Subsections 3.3 and 3.4.

\subsection{Unbroken N=(1,0) supersymmetry}
In the WZ gauge, unbroken supersymmetry acts via
\be
\delta_\eps V^{++}_{\W\Z} = (\eps^{+\a}\pa_{+\a}+\eps^{-\a}\pa_{-\a})
V^{++}_{\W\Z} - D^{++}\Lambda_\eps - [V^{++}_{\W\Z}, \Lambda_\eps]_\star
\label{27}
\ee
where  the standard piece contains $\eps^{\pm\a} = 
\eps^{\a k} u^{\pm}_k$, while the second piece is the
compensating gauge transformation with the superparameter 
(in the analytic basis, for brevity)
\bea
\Lambda_\eps(\zeta,u) &=& 2 (\epsilon^-\theta^+) \bph +
2 (\epsilon^-\sigma_m \bar\theta^+) A_m +  2(\bar\theta^+)^2
(\epsilon^-\Psi^-) \nn
&&+\,4(\epsilon^-\theta^+)(\bar\theta^+\bar\Psi^-)
+2(\epsilon^-\theta^+)(\bar\theta^+)^2{\cal D}^{--}\,.
\lb{28}
\eea
Note that the transformation law of $V^{--}$ has the same form \p{27} with
$(D^{++},V^{++}_{\W\Z}) \rightarrow (D^{--},V^{--})$. Also note that the
structure of $\Lambda_\eps$ is such that
only a term $\sim I$ is present in the commutator term in \p{27}.

The deformed $N{=}(1,0)$ supersymmetry transformation laws can be easily 
read off from the superfield transformation \p{27} as
\bea
&&\delta_\epsilon\phi=2(\epsilon^k \Psi_{k})\,,\q\delta_\epsilon\bph=0\,,\q
\delta_\epsilon A_m=(\epsilon^{k}\sigma_m \bar\Psi_k) \,,\nn
&&\delta_\epsilon\Psi^k_\a=-\epsilon_{\a l}{\cal D}^{kl}
+\frac12(1{+}4I\bph)(\si_{mn}\eps^k)_\a
F_{mn}-4\ii I\epsilon^k_\a A_m\pa_m\bph\,,\nn
&&\delta_\epsilon\bar\Psi^k_\da=
-\ii(1{+}4I\bph)(\epsilon^{k}\sigma_m)_\da\pa_m\bph\,,\nonumber\\[4pt]
&& \delta_\epsilon {\cal D}^{kl}=
\ii\pa_m[(\epsilon^k\si_m\bar\Psi^l+\epsilon^l\si_m\bar\Psi^k)
(1{+}4I\bph)]\,.\lb{defsusy}
\eea

\subsection{Invariant action}
The non-polynomial superfield action of the Q-deformed gauge theory has been
written in \cite{ILZ} as an integral over the full $N{=}(1,1)$ superspace
in chiral coordinates. In the QS-deformed U(1) theory which we consider here
this form of the action can be derived from the well-known formula
for its variation \cite{GIKOS}
\be
\delta S \sim \int \!\diff^4x_\sL \diff^4\th \diff^4\bt \diff u\  
\delta V^{++}\star\,V^{--}=
\int \!\diff^4x_\sL \diff^4\th \diff^4\bt \diff u\ 
\delta V^{++}\,V^{--}.\lb{15}
\ee
An equivalent form of the action can be constructed in chiral superspace,
\be
S= \frac14 \int \!\diff^4x_\sL \diff^4\theta \diff u\  \cW\star\cW =
\frac14 \int \!\diff^4x_\sL \diff^4\theta \diff u\  {\cal W}^2\, \lb{151}
\ee
where the covariantly chiral superfield strength is defined by
\be
{\cal W} = -\frac14(\bar D^+)^2V^{--} \equiv {\cal A} 
+ \bar\theta^+_\da{\tau}^{-\da}+
(\bar\theta^+)^2 {\tau}^{-2} \lb{152}
\ee
and $\cA, \tau^{-\da}$ and $\tau^{-2}$ are some chiral superfields.
Note that $\cW$ transforms non-trivially under the residual gauge group
given by \p{resid} while $\cA$ is invariant,
\be
\delta_r \cW=[\cW,\L_a]_\star\,,\q \delta_r\cA=0\,.
\ee
The basic harmonic equation \p{hzc}, when applied to \p{152},
yields the following harmonic equations for $\cW$ and $\cA$,
\bea
&&D^{++}\cW+[V^{++}_{\W\Z},\cW]_\star=0,\nn
&&D^{++}\cA +2I\partial^\a_{+}v^{++}\pa_{-\a}\cA=
D^{++}\cA+4I\bph\,\tpa\pa_{-\a} \cA=0.\lb{harmW}
\eea

The proof of equivalence of the two forms of the action (as an integral 
over the full $N{=}(1,1)$ superspace or as one over its chiral subspace) 
amounts to showing that the variation of the action is the same in the 
two representations.
This latter statement can be checked in several simple steps via
\bea
&\delta&\!\!\!\int\!\!\diff u \diff^4x \diff^4\th\ \cW^2=
2\int\!\!\diff u\diff^4x \diff^4\th \diff^4\bt\ (\btp)^2(\btm)^2 \cW\delta\cW
\sim\int\!\!\diff u\diff^4x\diff^4\th\diff^4\bt\ (\btp)^2 \cW\delta V^{--}\nn
&&\sim\int\!\!\diff u\diff^4x\diff^4\th\diff^4\bt\ (\btm)^2 \cW\delta V^{++}
\sim\int\!\!\diff u \diff^4x \diff^4\th \diff^4\bt\ V^{--}\delta V^{++}\,.
\eea
Here, the relation $(\tp)^2=\Dp(\tp\tm)$ as well as \p{hzc} and \p{harmW} were
used, along with integration by parts with respect to a harmonic derivative.

Actually, it can be shown that only the component ${\cal A}$ contributes 
to \p{151} (see below) 
and that the harmonic integral in \p{151}  can be omitted (this
directly follows from the harmonic constraint \p{harmW}). Hence, a 
convenient point of departure for
finding the invariant action is the following form of the latter,
\be
S= \frac14 \int \!\diff^4x_\sL \diff^4\theta \diff u\  {\cal A}^2=
\frac14 \int \!\diff^4x_\sL \diff^4\theta\  {\cal A}^2\,.
\lb{153}
\ee

The basic problem one should solve in order to find the explicit form 
of ${\cal A}$ and
hence that of the action \p{153} consists in calculating 
$V^{--}$ from \p{hzc}
specialized to the Abelian gauge group U(1) and to $V^{++}_{\W\Z}$ of 
eq.~\p{WZanal}:
\bea
&&D^{++}V^{--}-D^{--}V^{++}_{\W\Z}+2I\,
(\partial^\a_+ V^{++}_{\W\Z} \partial_{-\a} V^{--}- 
\partial^\a_{-} V^{++}_{\W\Z} \partial_{+\a}V^{--})\nn
&&+\,\sfrac12 I^3\left[\pa^\a_-(\pa_+)^2V^{++}_{\W\Z} \pa_{+\a}(\pa_-)^2V^{--}
-\pa^\a_{+}(\pa_-)^2V^{++}_{\W\Z}  \pa_{-\a}(\pa_+)^2V^{--}\right] = 0\,. 
\lb{hzc1}
\eea

We work in the chiral basis, so it will be convenient to expand all quantities
over $\bar \theta$:
\bea \label{Vmexp}
V^{--}(Z_\C,u) &=& {v}^{--} + \bar \theta^+_{\dot \alpha}\, 
{v}^{-3\dot\alpha} +
\bar \theta^-_{\dot \alpha}\, {v}^{-\dot\alpha} +  (\bar \theta^+)^2\,{v}^{-4}
+(\bar \theta^-)^2\,\cA + (\bar \theta^+ \bar \theta^-)\, {\varphi}^{-2} \\
&& + \,(\bar\theta^+ \tilde{\sigma}^{mn}\bar\theta^-)\, {\varphi}^{-2}_{mn}
+ (\bar\theta^-)^2\bar \theta^+_{\dot \alpha}\, {\tau}^{-\dot\alpha} +
(\bar\theta^+)^2 \bar\theta^-_{\dot \alpha}\, 
{\tau}^{-3\dot\alpha} +(\bar\theta^+)^2(\bar\theta^-)^2
\,\tau^{-2} \nonumber
\eea
with $Z_\C=(x^m_\sL,\th^{\pm\a},\bt^{\pm\da})$.
Note that the chiral superfield ${\cal A}$ given in \p{152} and\p{153}
is just one of the coefficients in this expansion, so it can be obtained
as a solution of the proper chiral harmonic equations in the expansion
of \p{hzc1} over $\bar\theta^+_{\dot\alpha}, \bar\theta^-_{\dot\alpha}$.
It is rather easy to solve all such chiral equations, but we consider only
those which are actually needed for determining ${\cal A}$.
These equations arise as coefficients of the monomials $1, 
\bar\theta^-_{\dot\alpha},
\bar\theta^+_{\dot\alpha}, (\bar\theta^-)^2$ and $(\bar\theta^+\bar\theta^-)$,
respectively:
\bea
&& \mbox{(a)}\quad \nabla^{++} v^{--}-D^{--}v^{++} = 0\,, \nn
&& \mbox{(b)}\quad \nabla^{++} v^{-\dot\alpha}-v^{+\da} = 0\,,\nn
&&\mbox{(c)}\quad \nabla^{++} v^{-3\da}-D^{--} v^{+\da}
+[v^{+\da},v^{--}]_\star =0\,,\nn
&& \mbox{(d)}\quad \nabla^{++} \cA = 0\,, \nn
&& \mbox{(e)}\quad \nabla^{++} \varphi^{-2} + 2 \cA - 2v
+ \sfrac12\{v^{+\da},  v^-_\da\}_\star  = 0
\lb{f}
\eea
where the star products are defined in \p{star}, \p{12} and the chiral U(1)
covariant harmonic derivative is defined as
\be
\nabla^{++}= D^{++}+[v^{++},~~]_\star = \pa^{++}
+ (1{+}4I\bph)\,\theta^{+\alpha}\partial_{-\alpha}\,. \label{defnabla}
\ee
The solutions of \p{f} contain positive and negative powers of the
function $1{+}4I\bph$.

We firstly address (\ref{f}a), (\ref{f}b) and (\ref{f}c). 
Their general solution is
\bea
&& v^{--} = (\theta^-)^2 \frac{\bar\phi}{1{+}4I\bar\phi}\,, \quad
v^{-3\dot\alpha} = 2(\theta^-)^2 \frac{1}{(1{+}4I\bar\phi)^2}\,
\bar\Psi^{-\dot\alpha}\,, \nn
&& v^{-\dot\alpha} = 
-2(\theta^-)^2 \frac{1}{(1{+}4I\bar\phi)^2}\,\bar\Psi^{+\dot\alpha}
+ 2(\theta^-\sigma_m)^{\dot\alpha}\frac{1}{1{+}4I\bar\phi} A_m  \nn
&&\qquad\quad
-\,4(\theta^+\theta^-)\frac{1}{1{+}4I\bar\phi}\,\bar\Psi^{-\dot\alpha}
+ 2\ii\,(\theta^-)^2 (\theta^+\sigma_m)^{\dot\alpha}
\frac{1}{1{+}4I\bar\phi}\,\partial_m \bar\phi\,.\label{abcsolv}
\eea
Now we are prepared to solve the remaining harmonic equations 
(\ref{f}d) and (\ref{f}e).
Using the explicit expression for $v^{-\dot\alpha}$ in \p{abcsolv},
these equations can be written in the unfolded form as
\bea
&& [\pa^{++}+ (1{+}4I\bar\phi)\tpa\pa_{-\a}]\cA=0\,, \lb{homA}\\[4pt]
&&[\pa^{++}+ (1{+}4I\bar\phi)\tpa\pa_{-\a}]\vp^\m
+2(\cA-v)
 -I\left(\pa^\a_{-}v^+_\da\pa_{+\a}v^{-\da}
-\pa_+^\a v^+_\da\pa_{-\a}v^{-\da}\right)
\nn
&&\qquad\qquad\qquad\qquad\qquad\qquad
+ \,\sfrac{1}{4}I^3\,\pa^\a_-(\pa_+)^2v^+_\da\pa_{+\a}(\pa_-)^2v^{-\da}
=0.\lb{inhom}
\eea
Then a straightforward computation yields the following rather 
lengthy solution for ${\cal A}$:
\bea
{\cal A}(z_c, u) &=& \left[\phi +4I\,(A_m)^2\,\frac{1}{1{+}4I\bar\phi}  + 
16 I^3 (\partial_m
 \bar\phi)^2
\,\frac{1}{1{+}4I\bar\phi}\right] \nn
&& +\, 2\theta^+ \left[\Psi^- 
+\frac{4I}{1{+}4I\bar\phi}(\sigma_m \bar\Psi^-)\,A_m \right]
- \frac{2}{1{+}4I\bar\phi}\theta^- \left[\Psi^+ 
+ \frac{4I}{1{+}4I\bar\phi}(\sigma_m \bar\Psi^+)\,A_m \right] \nn
&& +\, (\theta^+)^2 \left[\frac{8I}{1{+}4I\bar\phi}(\bar\Psi^-)^2 
+ {\cal D}^{--}\right] \nn
&& +\, \frac{1}{(1{+}4I\bar\phi)^2} (\theta^-)^2
\left[\frac{8I}{1{+}4I\bar\phi}(\bar\Psi^+)^2 + {\cal D}^{++}\right] \nn
&& -\,\frac{2}{1 + 4I\bar\phi} (\theta^+\theta^-)
\left[\frac{8I}{1{+}4I\bar\phi}(\bar\Psi^+\bar\Psi^-) + {\cal D}^{+-}\right]\nn
&& +\, (\theta^+\sigma^{mn}\theta^-)
\left(F_{mn} -8I\,\partial_{[m}\bar\phi A_{n]}\frac{1}{1{+}4I\bar\phi} 
\right) \nn
&& +\, 2\ii(\theta^-)^2\,\theta^+\si_m 
\partial_m \left( \frac{1}{1{+}4I\bar\phi}\bar\Psi^+\right)
+ 2\ii (\theta^+)^2\,(1{+}4I\bar\phi)\,\theta^-\sigma_m 
\partial_m \left( \frac{1}{1{+}4I\bar\phi} \bar\Psi^-\right) \nn
&& -\,(\theta^+)^2(\theta^-)^2\,\Box \bar\phi \label{vfull}
\eea
where $\ {\cal D}^{++} = {\cal D}^{kl}u^+_ku^+_l\ $ and $\ \q 
{\cal D}^{+-} = {\cal D}^{kl}u^+_ku^-_l\, $.

We observe that the shifted fermions $\Psi^\pm$, 
the combination of $\phi$ and $(A_m)^2$ in 
the first term
and the shifted $F^{mn}$ coincide (sometimes modulo powers of the factor
$1 + 4I\bar\phi$)
with the fields having standard transformation rules under the 
residual gauge group, as
they were introduced in \p{14}. In fact, \p{vfull} prompts the 
further canonical field redefinition
\bea
\vp&=&(1{+}4I\bar\phi)^{-2}[\phi+4I (1{+}4I\bar\phi)^{-1}
[(A_m)^2+4I^2(\pa_m\bph)^2]]\,,
\nn
a_m&=& (1{+}4I\bar\phi)^{-1}A_m,\q \bar\psi^k_\da=
(1{+}4I\bar\phi)^{-1}\bJ^k_\da\,,\nn
\psi^k_\a&=&(1{+}4I\bar\phi)^{-2}[\Psi^k_\a+4I(1{+}4I\bph)^{-1}
A_\ada\bJ^{\da k}]\,,\nn
d^{kl}&=&(1{+}4I\bar\phi)^{-2}[{\cal D}^{kl}+
8I(1{+}4I\bar\phi)^{-1}\bJ_\da^k\bJ^{\da l}]\,.
\lb{U1swrel}
\eea
It is straightforward to check that the deformed $N{=}(1,0)$ 
supersymmetry \p{defsusy}
has the {\it standard\/} realization on the newly introduced fields, namely
\bea
&&\delta_\epsilon\vp=2(\epsilon^{k}\psi_{k})\,,\q\delta_\epsilon\bph=0\,,\q
\delta_\epsilon a_m= (\epsilon^{k}\sigma_m \bar\psi_k)\,,\nn
&&\delta_\epsilon\psi^k_\a=-\epsilon_{\a l}d^{kl}+\sfrac12(\si_{mn}\eps^k)_\a
f_{mn}\,,\;\; \delta_\epsilon\bar\psi^k_\da=
-\ii(\epsilon^{k}\sigma_m)_\da\pa_m\bph\,,\nn
&&\delta_\epsilon d^{kl}=\ii\pa_m (\epsilon^k\si_m\bar\psi^l
+\epsilon^l\si_m\bar\psi^k)\,.
\lb{normsusy}
\eea
It is worthwhile to note that the appearance of the invariant term 
$16I^3(\pa_m\bph)^2(1{+}4I\bph)^{-3}$ in $\vp$ in the definition 
(\ref{U1swrel}) of $\vp$ is not suggested by our symmetry considerations.
However, the presence of this term in the field redefinition will simplify 
the final form of the invariant action.

Using the explicit expression \p{vfull} for ${\cal A}$, it is very easy 
to find the precise component expression of the QS-deformed superfield 
action \p{153} in terms of the fields introduced in \p{U1swrel}. The result is
\be
S =\frac14 \int \!\diff^4x_\sL \diff^4\theta\ {\cal A}^2 = 
\int \!\diff^4x\ L = \int \!\diff^4x\ (1 +4I\bar\phi)^2 L_0 \lb{ActionC}
\ee
where
\bea
&&L_0=-\frac12\vp\Box\bph+\frac14(f_{mn}^2
+\sfrac12\varepsilon_{mnrs}f_{mn}f_{rs})
-\ii\psi^\a_k\pada\bar\psi^{\da k}+\frac14(d^{kl})^2 \lb{LagrC}
\eea
is the standard undeformed Lagrangian of $N{=}(1,1)$ supersymmetric 
U(1) gauge theory. Thus we see
that the only effect of the nilpotent QS-deformation 
of the Abelian $N{=}(1,1)$ gauge theory is
the appearance of a scalar factor in front of the undeformed Lagrangian. 
This factor breaks half of the original supersymmetry and is in fact 
unremovable. Indeed, by properly rescaling all the involved
fields except for the gauge ones one can remove this factor from all pieces 
of the action \p{ActionC}, \p{LagrC} except from the gauge fields term
\be
\sfrac14(1{+}4I\bph)^2(f_{mn}^2+f_{mn}\tilde{f}_{mn})\,, \quad \tilde{f}_{mn}
\equiv \sfrac12\varepsilon_{mnrs}f_{rs}\,.
\ee
This nonlinear interaction of the fields $\bph$ and $f_{mn}$ cannot be removed
by any redefinition of fields.

As a final remark, let us explain why the chiral 
superfields $\tau^{-\da}$ and $\tau^{-2}$
in \p{152} do not contribute to the action \p{151}, 
which is therefore given by \p{153}.
The harmonic zero-curvature condition \p{hzc1} implies the equation 
\be
\nabla^{++}\tau^{-\da} + [v^{+\da}, \cA]_\star = 0\,,
\ee
which is solved by
\be
\tau^{-\da} = [\cA, v^{-\da}]_\star\,.
\ee
Then one finds that one of two terms with $\tau^{-\da}$ in \p{151} 
is vanishing:
\be
2\int \!\diff^{4}\theta\ \cA\star \tau^{-\da} = 
\int \!\diff^4\theta\ [(\cA\star\cA),
v^{-\da}] = 0\,.
\ee
Analogously, inspecting the harmonic equation for $\tau^{-2}$, 
one can check that the combined term with both $\tau^{-\da}$ and 
$\tau^{-2}$ in \p{151} is zero as well.

\subsection{Superfield Seiberg-Witten transform}
The field redefinition \p{U1swrel} (Seiberg-Witten map) relating fields 
with the deformed and undeformed U(1) gauge and $N{=}(1,0)$ supersymmetry 
transformation properties can be given a nice superfield form. Namely, 
it can be rewritten
as a relation between the covariantly chiral and covariantly 
harmonic-independent
superfield strength $\cA$ satisfying the constraint \p{harmW} and the
undeformed chiral U(1) superfield strength
\bea
W_0 (x_\sL, \theta^+, \theta^-,u)&=&\varphi+2\theta^+\psi^- -2\th^-\psi^+
+(\tp)^2 d^{--}\nn
&&-\,2(\tp\tm)d^{+-}+(\tm)^2d^{++}+
(\tm\sigma_{mn}\tp) f_{mn} \lb{undef}\\
&&+\,2\ii[(\tm)^2\th^+\sigma_m \pa_m\bar\psi^+ 
+(\tp)^2\th^-\si_m\pa_m\bar\psi^{-}]
-(\tp)^2(\tm)^2\Box\bph \nonumber
\eea
(where $\psi^\pm_\a = \psi^i_\a(x_\sL) u^\pm_i,~ d^{+-}=u^+_ku^-_ld^{kl} $
etc.), which obeys the free harmonic equation
\be
D^{++}W_0 = 0 \lb{freeconstrW}
\ee
and transforms under $N{=}(1,0)$ supersymmetry \p{normsusy} as
\be
\delta_\epsilon W_0 = 
(\epsilon^{-\alpha}\pa_{-\a}+\epsilon^{+\a}\pa_{+\a})W_0\,. \lb{standrule}
\ee
This superfield SW transform is given by
\be
{\cal A}(x_\sL, \theta^+, \theta^-,u) = 
(1{+}4I\bar\phi)^2\; W_0 (x_\sL, \theta^+, 
(1{+}4I\bph)^{-1}\theta^-,u)
\;.\lb{293}
\ee
Indeed, taking into account the relations \p{U1swrel}, it is easy to check that
the components of $\cA$ defined by \p{293} are expressed through the original
deformed component fields just according to \p{vfull}. 
Also, it is not hard to see that
\bea
&&\nabla^{++}{\cal A}(x_\sL, \theta^+, \theta^-,u) =
(1{+}4I\bph)^2\left(\partial^{++} + \theta^{+\a}\partial{\;}'{}_{-a}\right)
W_0(x_\sL, \theta^+, \theta^{-}{}', u)=0 \nn
&& \textrm{with}\qquad
\theta^{-\a}{}'\equiv (1{+}4I\bar\phi)^{-1}\theta^{-\a}\,,
\eea
whence it follows that the constraints \p{harmW} and \p{freeconstrW} 
are equivalent to each other.

The relation \p{293} can be cast in the convenient operator form
\bea
\cA&=&[(1{+}4I\bph)^2-4I\bph(1{+}4I\bph)(\tm \pa_-)-4I^2\bph^2(\tm)^2(\pa_-)^2]
W_0,\nn
\pa_{-\a}\cA&=&[(1{+}4I\bph)\pa_{-\a}+2I\tm_\a\bph(\pa_-)^2]W_0\,. \lb{ansatz}
\eea
Using \p{ansatz}, it is easy to deduce the transformation law 
of ${\cal A}$ under $N{=}(1,0)$
supersymmetry as a consequence of the standard transformation rule 
\p{standrule} of $W_0$, 
\bea
\delta_\epsilon \cA&=&[(1{+}4I\bph)^2-4I\bph(1{+}4I\bph)(\tm \pa_-)-4I^2\bph^2
(\tm)^2(\pa_-)^2]\delta_\epsilon W_0\nn
&=&[(1{+}4I\bph)\epsilon^{-\a}\pa_{-\a} +\epsilon^{+\a}\pa_{+\a}]\cA\,.
\eea
This is another form of the $N{=}(1,0)$ supersymmetry transformation induced for
${\cal A}$ by the corresponding transformation of the harmonic gauge 
potential $V^{++}$ in the WZ gauge \p{27},
\bea
&&\delta_\epsilon \cA=(\eps^{-\a}\pa_{-\a}+\eps^{+\a}\pa_{+\a})\cA+
[\cA,\lambda_\eps]_\star\,.
\eea
Here, $\lambda_\eps=2 (\epsilon^-\theta^+) \bph $ is just the chiral part 
of the compensating gauge transformation parameter $\L_\eps$ given in~\p{28}.

Using \p{293} or \p{ansatz}, the deformed U(1) superfield gauge action \p{153}
can be expressed in terms of the Abelian undeformed object $W_0$ 
(modulo a total spinor derivative in the integrand, see the Appendix),
\be
S = \frac14\int \!\diff^4x_\sL \diff^4 \theta \ {\cal A}^2=
\frac14\int \!\diff^4x_\sL \diff^4 \theta \ (1{+}4I\bph)^2W^2_0.\lb{U1rel}
\ee
This also yields the component action \p{ActionC}, \p{LagrC} directly.
The chiral superfield action of the deformed U(1) gauge theory can readily
be recast as an integral over the full superspace, by
using the expression of $W_0$ via the undeformed harmonic connection $\Vm_0$. 

\setcounter{equation}{0}
\section{QS-deformation of non-Abelian gauge theory}
\subsection{Gauge and N=(1,0) supersymmetry transformations}
We shall use the deformed U($n$) matrix potential $\Vp$ and gauge parameter
$\L$, which for the example of U(2) read
\be
\Vp=\Vp_0+\sha\Vp_b\tau_b,\q \L=\L_0+\sha\L_b\tau_b
\ee
where $\tau_b$ with $b=1,2,3$ are the Pauli matrices. 
The component decomposition
for the matrix superfields $\Vp_{\W\Z}$  can be analyzed in analogy
with the U(1) case \p{WZchir}.
Once again, the WZ form of the U($n$) gauge transformation of the potential
is specified by the residual analytic matrix parameter
\bea
&&\L_r=\ii a(x_\A)=[\ii+2(\tm\si_m\btp)\pa_m-\ii(\tm)^2(\btp)^2\Box]a(x_\sL),
\nn
&&\Dp\L_r=2(\tp\si_m\btp)\pa_m a-2\ii(\tp\tm)(\btp)^2\Box a,\q\pa_{+\a}\L_r=0,
\nn
&&\pa_{-\a}\L_r= 2(\si_m\btp)_\a\pa_m a-2\ii\tm_\a(\btp)^2\Box a,\q
(\pa_-)^2\L_r=4\ii(\btp)^2\Box a
\eea
and reads
\bea
\delta V^{++}_{\W\Z} = D^{++}\Lambda_r + [V^{++}_{\W\Z},\Lambda_r] +
I\,[\partial^\alpha_+ V^{++}_{\W\Z}, \partial_{-\alpha}\Lambda_r]
-\sfrac{1}{4}I^2[(\partial_+)^2V^{++}_{\W\Z},(\partial_{-})^2\Lambda_r]\,. 
\lb{000}
\eea
The corresponding deformed component gauge transformations have the
following form:
\bea
&& \d_r\bph=-\ii[a,\bph],\q 
\d_r\bJ^k_\da=-\ii[a,\bJ^k_\da],\q
\d_r {\cal D}^{kl}= -\ii[a,{\cal D}^{kl}]\,,\nn
&&\d_r A_m=\hat\nabla_m a +2I\{\bph,\pa_m a\}\,,\nn
&&\d_r\phi=-\ii[a,\phi] -4I\{A_m,\pa_m a\}-4\ii I^2[\Box a,\bph]\,,\nn
&&\d_r\J^k_\a=-\ii[a,\J^k_\a]
-2I(\si_m)_{\a\dot\a}\{\bJ^{\dot\a k},\pa_m a\} \lb{Ungauge}
\eea
where
\be
\hat\nabla_m = \pa_m + \ii[A_m, \;]\;.
\ee
The necessary presence of anticommutators in \p{Ungauge} is not compatible
with the standard tracelessness conditions for the deformed matrix fields
and so implies that the gauge group should contain an Abelian U(1) factor.
For this reason we choose U(n) as the natural gauge group. Nevertheless, 
as will be argued below, the deformed U($n$) theory can be covariantly
reduced to a SU($n$) one, at the cost of a nonlinear elimination of the trace
(i.e. U(1)) parts of the involved fields in terms of the traceless SU($n$)
parts.

Like in the Abelian case, the $N{=}(1,0)$ supersymmetry transformations 
of fields in WZ gauge can be found from
\p{27} with the same matrix compensating parameter \p{28}. It is easy to 
show that the deformed matrix commutator term contains terms of the 
orders $1$, $I$ and $I^2$:
\bea
[\Vp,\Lambda_\eps]_\star&=& 
[\Vp,\Lambda_\eps] - I[\pa^\a_{+}\L_\eps,\pa_{-\a} \Vp]
+I[\pa^\a_-\L_\eps,\pa_{+\a}\Vp]\nn
&&-\sfrac14 I^2[(\pa_-)^2\L_\eps,(\pa_+)^2\Vp]
+\sfrac14 I^2[\pa^\b_{+} \pa^\a_-\L_\eps,\pa_{+\a} \pa_{-\b}\Vp]
\eea
where the condition $(\pa_+)^2\L_\eps=0$ was used.
The resulting  transformations of the deformed non-Abelian components
under $N{=}(1,0)$ supersymmetry can be obtained from these
relations:
\bea
&&\delta_\epsilon\phi=-2(\epsilon_k\Psi^k)~,\q
\delta_\epsilon \bar\phi=0~,\q \delta_\epsilon A_m=-(\eps_k\si_m\bJ^k)\,,\nn
&&\delta_\epsilon\Psi^k_\alpha=-\epsilon_{\alpha l}{\cal D}^{kl}
-\sfrac12\eps^k_\a[\phi,\bph]+\sfrac12(\sigma_{mn}\eps^k)_\a{\cal F}_{mn}\nn
&&\qquad +\,I\left[(\sigma_{mn}\eps^k)_\a\{\bph, F_{mn}\} 
- 2\ii\eps^k_\a\{A_m, \pa_m\bph\}\right]\nn
&&\qquad-\,2I^2\left(\eps^k_\a[\bph, \Box \bph] + 
\ii(\si_{mn}\eps^k)_\a[\pa_m\bph, \pa_n\bph]\right), \nn
&&\delta_\epsilon {\cal D}^{kl} = 
2\ii\eps^{(k}\si_m\hat\nabla_m\bar\Psi{}^{l)} + 
2 \eps^{(k}[\Psi^{l)},\bph]
+4\ii I\eps^{(k}\si_m\pa_m\{\bar\Psi^{l)}, \bph\}\,,\nn
&&\delta_\epsilon\bar\Psi^k_\da=-\ii(\eps^{k}\si_m)_{\dot\a}\hat\nabla_m \bph
-2\ii I(\eps^{k}\si_m)_{\dot\a}\{\bph,\pa_m\bph\} \lb{nonabS}
\eea
where
\be
{\cal F}_{mn} = F_{mn} + \ii[A_m, A_n]\,.
\ee

\subsection{Basic chiral superfield  and invariant action}
We start with the covariantly chiral superfield strength
of the undeformed U($n$) gauge theory in the harmonic $N{=}(1,1)$ superspace
\cite{GIOS},
\be
W(Z_\C,u)=-\frac14(\bar D^+)^2V^{--}=A+\btp_\da \tau^{-\da}+(\btp)^2
\tau^{-2}
\lb{covch}
\ee
where $V^{--}$ is the undeformed harmonic connection and $A$, $\tau^{-\da}$
and $\tau^{-2}$ are the corresponding matrix chiral superfunctions living on
the subspace $(z_\C, u) = (x_\sL, \theta^\pm, u)$. The undeformed
U($n$) gauge theory action can be constructed in the chiral superspace as
\be
S_n=\frac14\int \!\diff^4x_\sL \diff^4\th\ \Tr W^2
=\frac14\int \!\diff^4x_\sL \diff^4\th\ \Tr A^2\,.
\ee
The undeformed harmonic chiral U($n$) superfield $A$
has the component expansion
\bea
A &=& \varphi+2\th^+ \psi^- -2\th^- \psi^+ +(\tp)^2 d^{--}
+(\tp\tm)([\vp,\bph]-2d^{+-})+(\tm)^2d^{++} \nn
&& +\, (\tp\si_{mn}\tm) f_{mn}+2(\tm)^2\th^+\left(\ii\xi^+ -[\bph,\psi^+]
\right)+2\ii(\tp)^2\th^-\xi^- \nn
&& -\,(\tp)^2(\tm)^2\left(p+[\bph,d^{+-}]\right)\lb{Unundef}
\eea
where all component fields are $n{\times}n$ matrices 
and the following short-hand
notation is used:
\bea
&&\nabla_m=\pa_m +\ii[a_m,~~~]\,,\q 
f_{mn}=\pa_ma_n-\pa_na_m+\ii[a_m,a_n]\,,\nn
&&\xi^k_\a=(\si_m)_\ada\na_m\bj^{\da k},\q p
=\nabla_m^2\bph+\{\bar\psi^{\da k},\bar\psi_{\da k}\}
+\sfrac12[\bph,[\bph,\vp]]\,.\lb{Acomp}
\eea
Note that last two terms in $p$ are traceless, i.e. 
belong to the $su(n)$ algebra. So $p$ can
be decomposed into trace and traceless parts as 
\be
p = p_0 + \hat{p} = \Box \bph_0 + \nabla_m^2\hat{\bph}
+\{\bar\psi^{\da k},\bar\psi_{\da k}\}
+\sfrac12[\bph,[\bph,\vp]]\,,\quad  \Tr\, \hat{p} =\Tr\, \hat{\bph} =0\,.
\lb{tracel}
\ee

In the WZ gauge, the harmonic chiral superfield $A$ satisfies
the condition
\be
\Dp A + [v^{++}, A] = \Dp A + (\tp)^2[\bph,A] = 0\,.
\ee
The undeformed $N{=}(1,0)$ supersymmetry transformation of 
this chiral superfield is
\be
\d_\eps A=(\eps^{-\a}\pa_{-\a}+\eps^{+\a}\pa_{+\a})A+2(\eps^-\tp)[\bph,A]\,.
\ee
The corresponding component transformations are written down in the Appendix.

The harmonic zero-curvature equation for the {\it deformed\/} U($n$) harmonic
connections \p{hzc} contains terms of all orders in the deformation 
parameter, i.e.
$\sim 1, I, I^2, I^3, I^4$. It is convenient to use the chiral 
expansions \p{WZchir} and \p{Vmexp} for this case, since then one obtains 
the proper set of harmonic equations for the U($n$)-matrix chiral components. 
The most important equations of actual need for restoring the 
non-Abelian analog of the action \p{ActionC} are
\bea
&&\nabla^{++} v^-_\da -v^+_\da +I^2[\bph,(\pa_-)^2v^-_\da]=0\,,
\lb{vequa}\\[6pt]
&&\nabla^{++}\cA+I^2[\bph,(\pa_-)^2\cA]=0\,,\lb{harmcA}\\[6pt]
&&\nabla^{++}\vp^\m +2\cA-2v + \sfrac12 \{v^-_\da, v^{+\da}\}
+ \sfrac12 I \bigl[ \{\pa^\a_{+} v^{+}_{\da},\pa_{-\a}v^{-\da}\} 
-\{\pa^\a_- v^{+}_{\da},\pa_{+\a} v^{-\da}\}\bigr]\nn
&&\quad+\,I^2[\bph,(\pa_-)^2\vp^\m] -\sfrac18 I^2
\{(\pa_-)^2v^-_\da,(\pa_+)^2v^{+\da}\}
-\sfrac12 I^2\{\pa^\a_{+} \pa^\b_-v^-_\da, \pa_{+\b} \pa_{-\a}v^{+\da}\} \nn
&&\quad+\,\sfrac18 I^3\{\pa^\a_-(\pa_+)^2 v^+_\da,\pa_{+\a}(\pa_-)^2v^{-\da}\}
=0 \lb{inhomcA}
\eea
where
\be
\nabla^{++} = D^{++} +(\tp)^2[\bph, ~~~] + 
2I\theta^{+\alpha}\partial_{-\a}\{\bph, ~~~\}\,.
\ee

All these equations contain commutators and anticommutators 
with the  matrix field $\bph$. The
matrix solutions of these equations can be obtained by exploiting 
the method of
generating operators (functions) which allows one to explicitly calculate 
coefficients
in matrix series. The basic quantities of this method are operators 
$X$ and $Y$ which generate the left and right
matrix multiplication with $\bph$, 
\be
XM=\bph M,\q YM=M\bph,\q X^kY^lM=\bph^kM\bph^l,\q [X,Y]=0\lb{defXY}
\ee
where $M$ is some arbitrary matrix $([\bph,M]\neq 0)$ and $k, l$ are
positive integers. 
A linear inhomogeneous equation
\be
F(X,Y)\,M=B
\ee
for a matrix $M$, with a given matrix $B$ and some invertible
operator $F(X,Y)$ introducing a dependence on the matrix $\bph$
via \p{defXY}, can be solved by
\bea
&&M=F^{-1}(X,Y)\,B\,, \quad F^{-1}(X,Y)\,F(X,Y) = 1\,.\lb{phequ}
\eea

The homogeneous equation \p{harmcA} contains $\bph$-matrix
operators which can be represented in terms of the two generating
functions
\be
X-Y \quad  \mbox{and} \quad L(X,Y)=1+2I(X+Y)\,.
\ee
The solution for $\cA$ can be found in terms of powers of these operators
acting on the undeformed matrix component fields of the superfield
$A$ defined in \p{Unundef}. We obtain
\bea
\cA &=& L^2\vp-4I^2\hat{p}+4I^2 (X-Y)d^{+-}
+2\th^+\{[L^2-4I^2(X-Y)^2]\j^- \nn
&&+\,4\ii I^2(X-Y)\xi^-\}- 2\th^-L \psi^+ +(\tp\si_{mn}\tm) 
Lf_{mn}+(\tm)^2 d^{++} \nn
&&+\, (\tp)^2 [L^2-4I^2(X-Y)^2]d^{--} -(\tp\tm)[(X-Y)L\vp+2 Ld^{+-}]
 \nn
&& +\, 2(\tm)^2\tp [\ii\xi^+ -(X-Y)
\j^+] +2\ii(\tp)^2\tm L\xi^- \nn
&&-\,(\tp)^2(\tm)^2[p+(X-Y)d^{+-}]
\lb{UndefcA}
\eea
where $\xi^k_\a$, $p$ and $\hat{p}$ are the composite components of $A$
defined in \p{Acomp} and \p{tracel}.
Our short-hand notation for the matrix operators expands as
\bea
&&L^2\vp=\vp+2I\{\bph,\vp\}+4I^2\{\bph,\{\bph,\vp\}\},\q (X-Y)\vp
=[\bph,\vp]\,,\nn
&&Lf_{mn}=f_{mn}+2I\{\bph,f_{mn}\},\q (X-Y)^2\j^k_\a
=[\bph,[\bph,\j^k_\a]]\,.
\eea

It is worth noting that solutions of the homogeneous equation \p{harmcA}
are defined up to an arbitrary invariant operator multiplier $N(\bph)$.
In order to uniquely fix this normalization matrix one can make use of the
inhomogeneous equation \p{inhomcA} and one of the relations between 
the deformed and
undeformed U($n$) component fields. For two types of component fields
these relations can be directly deduced from the $N{=}(1,0)$ supersymmetry
transformation law \p{nonabS} and have the especially simple form
\be
\mbox{(a)}\quad A_m=La_m ,\qq \mbox{(b)} \quad \bar\Psi^k_\da
=L\bj^k_\da\,.\lb{simrel}
\ee
Substituting the inverse relation $a_m=L^{-1}A_m$ into our solution
\p{UndefcA} (corresponding to the choice $N(\bph)=1$), one can express
the $\tp\si_{mn}\tm$ component in $\cA$ via the deformed fields
\be
Lf_{mn}=L(\pa_ma_n-\pa_na_m+\ii[a_m,a_n])=\pa_mA_n-\pa_nA_m+\ldots
\ee
where the omitted higher-order terms contain no derivatives on $A_m$.
Just this normalization of the  term $\tp\si_{mn}\tm(\pa_mA_n-\pa_nA_m)$ 
in $\cA$ is compatible with the inhomogeneous equation \p{inhomcA}. 
Indeed, only
with its choice this term is canceled in \p{inhomcA} by the same 
structure coming
from the superfield $v$ defined in \p{v++}. Thus the choice $N(\bph)=1$ is
the correct one.

The $N{=}(1,0)$ supersymmetry transformation of $\cA$ can be obtained
from the superfield transformation \p{27} (which has the same form
both for the Abelian and non-Abelian cases),
\bea
\d_\eps\cA&=&(\eps^{-\a}\pa_{-\a}+\eps^{+\a}\pa_{+\a})\cA+[\l_\eps,
\cA]_\star+I[\pa^\a_+\l_\eps,\pa_{-\a}\cA]\nn
&=&2(\epsilon^-\theta^+)[\bph,\cA]+L\eps^{-\a}\pa_{-\a}\cA
+\eps^{+\a}\pa_{+\a}\cA
\eea
where $\ \lambda_\eps{=}2 (\epsilon^-\theta^+) \bph\ $ and 
$\,\q\pa_{+\a}\l_\eps{=} 2\eps^-_\a\bph\ $ as follows from \p{28}. 
Note that this deformed
superfield transformation could equally be deduced from the undeformed
$N{=}(1,0)$ transformations of the component fields given in \p{10un}.

The deformed chiral U($n$) superfield \p{UndefcA} can be written as a sum
of two $N{=}(1,0)$ covariant objects, 
\bea
\cA(x_\sL,\tp,\tm,u)&=& \bigl[ L^2+L(1{-}L)(\tm\pa_-)
-\sfrac14 (1{-}L)^2(\tm)^2(\pa_-)^2\bigr]\, A(x_\sL,\tp,\tm,u)\nn
&&-\,4I^2 \hat{A}(x_\sL,\tp,u)
\lb{Unoper} 
\eea
where $A$ is the undeformed U($n$) superfield \p{Unundef}. 
The matrix operator $L$ in \p{Unoper} acts on all quantities standing
to the right. The second part $\hat{A}$ is a traceless  
chiral-analytic $N{=}(1,0)$ superfield,
\bea
&&\hat{A}(x_\sL,\tp,u)=\hat{p}- [\bph,d^{+-}]
+2\tp(\ii[\bph,\xi^-]-[\bph,[\bph,\j^-]])
+(\tp)^2[\bph,[\bph,d^{--}]],\nn
&&\d_\eps\hat{A}=\eps^{+\a}\pa_{+\a}\hat{A}
+2(\epsilon^-\theta^+)[\bph,\hat{A}]\,,\qq
\pa_{-\a}\hat{A}=0\,.
\eea

The relation \p{Unoper} constitutes a non-Abelian analog of the U(1) 
superfield SW map \p{293}. It is obvious that the WZ gauge 
transformation of the deformed chiral superfield $\cA$ defined by 
\p{UndefcA} is of the same form as the transformation of $A$, i.e.
$$
\delta_r \cA = -\ii[a(x_\sL), \cA]\,, \q \delta_r A = -\ii[a(x_\sL), A]\,.
$$  
Thus, like in the Abelian case, the map \p{Unoper} does not imply 
a redefinition of the parameters $a(x_\sL)$ of the residual gauge 
transformations of the component fields, in contrast to the 
standard bosonic SW map \cite{SW}.

As noticed above, the components of $\cA$ can be expressed through 
the original deformed fields
via the inhomogeneous equation \p{inhomcA}. 
The equation \p{vequa} for $v^{-\da}$ can also be solved easily.  
The explicit relations between the deformed
and undeformed U($n$) fields are much more complicated as compared 
to the U(1) case. They are rather intricate even in the simplest $U(2)$ 
case (see Subsection 4.3).

It is worth emphasizing that the tracelessness condition $\mbox{Tr} W{=}0$
for the undeformed covariantly chiral U($n$) superfield strength $W$  
defined in \p{covch} is covariant with respect to both gauge transformations
(which are the same for deformed and undeformed fields) and supersymmetry
transformations.
This constraint eliminates the undeformed U(1) gauge multiplet and 
so reduces the U($n$) gauge group to SU($n$). It amounts to the same condition
for the chiral superfield $A$, i.e. $\mbox{Tr} A{=}0$. The latter, via the
nonlinear map \p{Unoper}, effectively expresses the trace part 
of the deformed superfield $\cA$ in terms of the traceless deformed fields.
Thus the existence of the SW map in the non-Abelian case allows one to
nonlinearly reduce the deformed U($n$) gauge theory to a theory with
gauge group SU($n$), in which U(1) fields become composite.

The component action of the QS-deformed U($n$) gauge theory 
can be obtained from the superfield
chiral action by making use of the relations \p{UndefcA}, \p{Unoper}.
One arrives at
\be
{\cal S}_n=\frac14\int \!\diff^4x_\sL \diff^4\theta\ \Tr\cA^2=
\int \!\diff^4x_\sL \diff^4\theta\ \Tr\{\sfrac14 (LA)^2-2I^2\hat{A}A\}\,.
\lb{Sn} \ee
As follows from \p{harmcA} (and the explicit expression \p{UndefcA}
for ${\cal A}$), the harmonic-dependent terms of the superfield
$\Tr \cA^2$ cancel between themselves in the coefficient 
of $(\tp)^2(\tm)^2\,$.
The first term of the deformed U($n$) action ${\cal S}_n$ in (\ref{Sn})
contains the $L$-rescaled
undeformed superfields in complete analogy with the U(1) action \p{U1rel}:
\bea
\frac14\int \!\diff^4x_\sL \diff^4\theta\ \Tr (LA)^2 &=&
\int \!\diff^4x_\sL\ \Tr\Bigl[
\sfrac14(Lf_{mn})^2+\sfrac18\ve_{mnrs}Lf_{mn}Lf_{rs}-\sfrac12L\vp
L\na_m^2\bph\nn
&&\qquad\qquad+\,\sfrac14(L[\bph,\vp])^2
+\sfrac14(Ld^{kl})^2-\sfrac12L\vp L\{\bj^{\da k},\bj_{\da k}\} \nn
&&\qquad\qquad-\,\ii L\j^\a_kL\na_\ada\bj^{\da k}
+\sha \bph\{L\j_{\a k},L\j^{\a k}\}\Bigr].
\eea
The second term of ${\cal S}_n$ in (\ref{Sn}) has no analog in the U(1) case.
It contains  higher derivative terms for the traceless parts 
of the fields $\bph$ and $\bj^\da_k$,
\bea
-2I^2\int \!\diff^4x_\sL \diff^4\theta\ \Tr\,(\hat{A}A)&=&
2I^2\int \!\diff^4x_\sL\ \Tr\Bigl[
\hat{p}^2+\sha[\bph,d^{kl}]^2 +\bph\{\xi_{\a k},\xi^{\a k}\} \nn
&&\qquad-\,2[\bph,\j_{\a k}][\bph,\xi^{\a k}]+
\bph\bigl\{[\bph,\j_{\a k}],[\bph,\j^{\a k}]\bigr\}\Bigr] \lb{higher}
\eea
where
\bea
&&\hat{p}^2=(\na_m^2\hat{\bph})^2+\na_m^2\hat{\bph}[\bph,[\bph,\vp]]
+\sfrac14[\bph,[\bph,\vp]]^2
\nn &&\qquad+\,
\{\bj^{\da k},\bj_{\da k}\}^2+\{\bj^{\da k},\bj_{\da k}\}[\bph,[\bph,\vp]]
+2\na_m^2\hat{\bph}\{\bj^{\da k},\bj_{\da k}\},\lb{extra1} \\
&&\bph\{\xi_{\a k},\xi^{\a k}\}=\bph \ve^\ab\{\na_\ada\bj^\da_k,\na_\bdb
\bj^{\db k}\}\,.\lb{extra}
\eea

The appearance of higher-derivative interactions of $\bph$
and $\bj^{\da k}$ in \p{higher} can be attributed to the specific feature
of the relation between deformed and undeformed field variables in the 
non-Abelian case.
Namely, as follows from \p{UndefcA}, terms with higher derivatives are present
in this relation already at the linearized level. An example of such a term is
\be
-4I^2\Box \hat\bph +8\ii I^2\tpa u^-_k[\bph,\pada\bj^{\da k}]\,.
\ee
Being rewritten in terms of the initial deformed fields, the component action
will hopefully be of the correct order in derivatives (two on physical 
bosons and one on fermions).
We did not analyze this issue in more detail. The only statement 
we can be sure of is that no such
terms appear at the linearized level after field redefinitions.
Indeed, inspecting \p{higher}, \p{extra1} and \p{extra},
it is easy to observe that many higher-derivative terms can be absorbed into
a redefinition of the fields $\varphi$ and $\psi^k_\a$, 
leaving us with the correct free part for the entire action.

\subsection{Seiberg-Witten map for the gauge group U(2)}

It is interesting to obtain the explicit form of the SW map for the 
non-Abelian case.
It is easy to check that the transformations \p{Ungauge} have the same 
Lie-bracket structure as the standard U($n$) gauge transformations, 
so the existence of a transform to fields with conventional
gauge transformation properties is evident from the very beginning. 
While in the general case we are still not
aware of an explicit closed form for such a field redefinition, 
it can be easily found in
the particular case of the gauge group U(2). We will present firstly the 
non-Abelian analog of the ``minimal'' SW transform \p{14} for this case 
and secondly its modification which is an analog
of the field redefinition \p{U1swrel} and relates the original 
deformed fields to those possessing
undeformed gauge and $N{=}(1,0)$ supersymmetry transformation properties. 
The derivation below does not appeal
to any superfield considerations and uses as an input the component 
transformation rules \p{Ungauge} and \p{nonabS}.

We decompose
\be
A_m = A_m^0 {\bf 1} + \sfrac12{\bf A}_m\cdot  \boldsymbol{\tau}\,.
\ee
The deformed part of the gauge transformations \p{Ungauge} 
for this particular case amounts to
\bea
\delta A_m^0 &=& (1{+}4I\bar\phi^0)\partial_m a^0 
+ I\,(\boldsymbol{\bph}\cdot\partial_m {\bf a})\,, \nn
\delta {\bf A}_m &=& \nabla_m {\bf a}+
4I(\bph^0\partial_m {\bf a} + \boldsymbol{\bph}\partial_m a^0)\,, 
\label{u21} \\
\delta \phi^0 &=& -8I\,\bigl(A^0_m\partial_m a^0 
+\sfrac14{\bf A}_m \cdot \partial_m {\bf a}\bigr), \nn
\delta \tilde{\boldsymbol{\phi}} &=& 
-\tilde{\boldsymbol{\phi}}{\bf \times  a} - 8I\left({\bf A}\partial_m a^0 +
A^0_m\partial_m {\bf a}\right) + 
8I^2\,\partial_m \boldsymbol{\bph}{\bf \times}\partial_m {\bf a}\label{u22}
\eea
where
\be
\tilde{\boldsymbol{\phi}} = \boldsymbol{\phi} -4I^2 \,\Box \boldsymbol{\bph}\,.
\ee
The transformation law of $\Psi^k_\alpha$ can be easily cast 
in the standard form by the appropriate shift of $\Psi^k_\alpha$ 
(see \p{newpsi} below).

The ``minimal'' SW map (a non-Abelian analog of \p{14}) is given 
by the following relations,
\bea
A^0_m &=& \left(1{+}4I\,\bph^0\right) a^0_m + I\, 
(\boldsymbol{\bph}\cdot \boldsymbol{a}_m)\,, \quad
{\bf A}_m \ =\ \left(1{+}4I\,\bph^\0\right) {\bf a}_m
+ 4I \,\boldsymbol{\bph}\,a^0_m\,, \lb{Atoa}\\
\phi^0 &=& \hat{\phi}^0 -
I\left(1{+}4I\,\bph^\0\right)\left({\bf a}_m\cdot {\bf a}_m 
+ 4 a^0_m a^0_m \right)
- 8I^2\,({\bf a}_m\cdot\boldsymbol{\bph})\,a^0_m\,, \nn
\tilde{\boldsymbol{\phi}} &=& \hat{\boldsymbol{\phi}} 
-8I\left(1{+}4I\,\bph^0 \right) {\bf a}_m a^0_m -
4I^2\,\boldsymbol{\bph}\left({\bf a}_m\cdot{\bf a}_m 
+ 4 a^0_m a^0_m \right)
+\,8I^2\,\partial_m \boldsymbol{\bph}{\bf \times a}_m\,. 
\label{SWu2}
\eea
Here $a^0_m$, ${\bf a}_m$ are U(2) gauge fields and $\hat{\phi}^0, 
\hat{\boldsymbol{\phi}}$
are ``matter'' fields with the standard gauge transformation properties
\bea
\delta a^0_m = \partial_m a^0\,, \quad \delta {\bf a}_m =
\partial_m {\bf a} - {\bf a}_m {\bf \times a}\,, \quad
\delta \hat{\phi}^0 = 0\,, \quad \delta \hat{\boldsymbol{\phi}} =
-\hat{\boldsymbol{\phi}}{\bf \times a}\,. \label{stand1u2}
\eea
It is straightforward to check that \p{stand1u2}  generate 
just the transformations
\p{u21} and \p{u22} of the original fields. It is obvious that 
the map \p{SWu2} is invertible, but inverse relations look not 
too illuminating.

Taking for granted that an analogous ``minimal'' SW map exists 
in the general U($n$) case too,
the shifted field $\Psi^k_\alpha$ with the standard gauge 
transformation law is constructed as
\be
\hat{\Psi}^k_\alpha = \Psi^k_\alpha + 
2I(\sigma_m)_{\alpha\dot\alpha}\{\bar\Psi^{k\dot\alpha}, a_m\}\,, \quad
\delta a_m = \partial_m a + \ii[a_m, a]\,. \label{newpsi}
\ee

It remains to see
whether it is possible, as in the Abelian case, to combine the
``minimal'' SW transform \p{SWu2} with some additional field redefinition, 
so as to convert
\p{nonabS} into the standard form of undeformed $N{=}(1,0)$ supersymmetry.

It is straightforward to check that the fields $\bar\psi^{k\,0}_\da$ 
and $\boldsymbol{\bar\psi}^{k}_\da$
related to $\bar\Psi^{k\,0}_\da$ and $\boldsymbol{\bar\Psi}^{k}_\da$ by
an invertible relation similar to \p{Atoa},
\be
\bar\Psi^{k\,0}_\da = \left(1{+}4I\,\bph^0\right) \bar\psi^{k\,0}_\da +
I\, (\boldsymbol{\bph}\cdot \boldsymbol{\bar\psi}^{k}_\da)\,, \quad
\boldsymbol{\bar\Psi}^{k}_\da = 
\left(1{+}4I\,\bph^\0\right) \boldsymbol{\bar\psi}^{k}_\da
+ 4I \,\boldsymbol{\bph}\,\bar\psi^{k\,0}_\da\,, \lb{Atoa1}
\ee
undergo, together with $a_m^0, {\bf a}_m$, just the 
undeformed $N{=}(1,0)$ supersymmetry transformations
\be
\delta_\epsilon a_m=-\eps_k\si_m\bar\psi^k\,,\quad
\delta_\epsilon\bar\psi^k_\da=-\ii(\eps^{k}\si_m)_{\dot\a}(\pa_m \bph 
+ \ii[a_m, \bph])\,.
\ee
Note that the field redefinition \p{Atoa}, \p{Atoa1} is the particular 
U(2) case of the general U($n$) relations \p{simrel}.

In a similar direct way one can find the corresponding ``full'' 
SW maps for the remaining U(2) fields $\phi, \Psi^k_\a$ and 
${\cal D}^{kl}$ which transform under $N{=}(1,0)$ supersymmetry
according to the deformed laws \p{nonabS}. These relations look rather 
ugly even in the U(2) case. As an example we give here how the singlet 
auxiliary field $d^{kl\,0}$ with the standard $N{=}(1,0)$ transformation 
law is expressed through the deformed components as
\bea
d^{kl\,0} &=& \alpha(\bph) {\cal D}^{kl\,0} 
-2I \beta(\bph)\,(\boldsymbol{\bph}\cdot\boldsymbol{\cal D}^{kl})
+ 8\gamma(\bph)(\bar\psi^{k\,0}\bar\psi^{l\,0}) 
+ 2I \sigma(\bph)(\boldsymbol{\bar\psi}^k\cdot
\boldsymbol{\bar\psi}^l) \nn
&& -\,16I^2 \omega(\bph) (\bar\psi^{(k\,0}\boldsymbol{\bph}\cdot
\boldsymbol{\bar\psi}^{l)})
- 8 I^3 \beta(\bph)
(\boldsymbol{\bph}\cdot \boldsymbol{\bar\psi}^k)
(\boldsymbol{\bph}\cdot \boldsymbol{\bar\psi}^l)\,.
\eea
Here,  $\bar\psi^{k\,0}_\da$ and $\boldsymbol{\bar\psi}^{k}_\da$ 
should be expressed through
$\bar\Psi^{k\,0}_\da$ and $\boldsymbol{\bar\Psi}^{k}_\da$ 
from \p{Atoa1}, and the functions $\alpha, \beta,
\gamma, \sigma$ and $\omega$ are given by the expressions
\bea
&& \alpha = G_+ \,(G_-)^{-2}\,, \; \beta = g \,(G_-)^{-2}\,, \; 
\gamma = g \,(G_-)^{-1}\,, \;
\sigma = g^3 \,(G_-)^{-2}\,, \; \omega = (G_-)^{-1}\,,\nn
&& \textrm{with} \qquad
g \equiv 1 + 4I\,\bph^0 \qquad\textrm{and}\qquad 
G_{\pm} \equiv g^2 \pm 4I^2\,(\boldsymbol{\bph}\cdot \boldsymbol{\bph})\,. 
\lb{Aux}
\eea

In accordance with the discussion in Subsection 4.2, in the general U(2)
SW map one can covariantly set undeformed U(1) fields equal to zero,
thus reducing the gauge group U(2) to SU(2). Then the deformed U(1)
fields will become composite functions of the undeformed SU(2)  
fields  but of course they can be re-expressed as well through the deformed
SU(2) fields.

\setcounter{equation}{0}
\section{Relation to the string background}

In this section we comment on a possible interpretation of the superspace 
deformation discussed in this paper as arising in string theory with 
a specially chosen constant background, along the lines of \cite{OoVa,Se}.
We start by briefly recalling the $N{=}1$ case. There one chooses 
a background consisting
of a constant self-dual two-form $F^{\a\b}$ (the graviphoton field strength).
This choice is consistent with the background supergravity field 
equations since
the self-dual field strength does not contribute to the stress tensor 
and thus does not
create a back reaction of the metric. The coupling of an $N{=}2$ 
superstring propagating
in four space-time dimensions to this constant background leads 
to the effective Lagrangian
\begin{equation}\label{R1}
  {\cal L}_{\textrm{eff}} = \left(\frac{1}{{\a'}^2 F}\right)_{\a\b} 
\pa\tilde\theta^\a \bar\pa
  \theta^\b\,.
\end{equation}
Imposing the appropriate boundary conditions which halve the number 
of Grassmann variables
and solving for the $\theta$ propagators, one finds that they lead 
to the deformed anticommutator
\begin{equation}\label{R2}
  \{\theta^\a,\theta^\b\} = {\a'}^2 F^{\a\b}\,,
\end{equation}
where the constant tensor $F$ plays the r\^ole of the deformation parameter.

{}From the point of view of the compactification of the type IIB background
one may say\footnote{We are grateful to A. Lerda for a clarifying discussion 
of this point (see also \cite{Lerda}).}
that the self-dual tensor $F^{\a\b}$ originates from the Ramond-Ramond sector,
more precisely, from the five-form in ten dimensions wrapped around
the (3,0)-form of the Calabi-Yau manifold viewed as
an orbifold ${\mathbb C}^3/Z_2\times Z_2$. If we want
to find the origin of the singlet scalar deformation parameter 
discussed in this paper,
we should look in a different part of the Ramond-Ramond sector, 
namely at the one-form.
This time we should imagine a compactification on 
${\mathbb C}\times{\mathbb C}^2/Z_2$.
The one-form is the derivative of the axion field, $F_\mu = \pa_\mu C$.
After the compactification we restrict it to just the two dimensions
of the torus, $F_a = \pa_a C$, $a=4,5$, or in complex notation,
$F = \pa_\tau C$, $\tau = x^4 + \ii x^5$. Further, the kinetic term of the axion 
is of the form
\begin{equation}\label{R3}
  \pa_a C \pa_a C = \pa_\tau C \pa_{\bar\tau} C\,.
\end{equation}
Our (Euclidean) choice will be
\begin{equation}\label{R4}
  C = \tau F_0
\end{equation}
where $F_0$ is a real constant, i.e. $C$ will be a real analytic function.
Then, since $\pa_{\bar\tau} C=0$, we see that the kinetic term (\ref{R3}),
and with it the contribution to the stress tensor, vanishes. We conclude
that this is a consistent choice of the background.

The rest closely follows the $N{=}1$ case described above. 
We couple this background
to an $N{=}4$ superstring and obtain the  effective Lagrangian
\begin{equation}\label{R5}
  {\cal L}_{\textrm{eff}} = \left(\frac{1}{{\a'}^2 F_0}\right)
  \epsilon_{\a\b} \epsilon_{ij}\,\pa\tilde\theta^{i\a} \bar\pa \theta^{j\b}\,.
\end{equation}
The boundary conditions halve the number of $\theta$s, and the resulting
propagators lead to the deformed superspace  anticommutator
\begin{equation}\label{R6}
  \{\theta^{i\a},\theta^{j\b}\} = {\a'}^2 F_0 \epsilon^{\a\b} \epsilon^{ij}\,,
\end{equation}
where the constant $F_0$ plays the r\^ole of the deformation parameter.

\newpage

\section{Conclusion}
In this paper we have studied the structure of QS-deformed Euclidean 
$N{=}(1,1)$ gauge theories,
both in the Abelian and non-Abelian cases. This type of 
nilpotent Q-deformation
is distinguished in that it preserves the ``Lorentz'' 
O(4) symmetry and the SU(2)
R-symmetry. It breaks $N{=}(1,1)$ supersymmetry down to $N{=}(1,0)$ 
but preserves both chirality and
Grassmann harmonic analyticity. Using the harmonic superspace
approach, we have obtained the relevant non-linear invariant actions, 
both in the superfield
and component form and to any order in the deformation 
parameter $I$, as well as the deformed gauge and $N{=}(1,0)$ 
supersymmetry transformations which leave these actions invariant.
We have found that in the non-Abelian case the gauge group naturally
contains a U(1) factor, similarly to other non-commutative gauge field
theories. We have established the existence of a Seiberg-Witten-type
map from the fields with deformed gauge and supersymmetry transformations
to those having standard transformation properties. This map implies
that the deformed U($n$) gauge theory can be covariantly reduced
to the SU($n$) one, with the deformed U(1) fields becoming nonlinear
functions of the traceless SU($n$) fields. The explicit
form of the SW map has been given for the gauge groups U(1) and U(2).
We have also revealed the string theory origin of the QS-deformation
considered.

There are several possibilities for extending our results. 
Firstly, it is of obvious interest to
add the couplings to matter and to study the structure of the resulting 
scalar potential.
Secondly, it would be tempting to analyze general Q-deformed $N{=}(1,1)$ 
gauge theories (where both O(4) and R-symmetry SU(2) are broken) 
along similar lines and to investigate 
their possible relation to strings. Finally, an obvious and urgent task 
would be to study the quantum properties of the nilpotently deformed 
$N{=}(1,1)$ gauge theories constructed in \cite{ILZ, FS} and in the present 
paper.

\section*{Acknowledgements}
The work of S.F., E.I., E.S. and B.Z. has been supported in part 
by the INTAS grant No 00-00254.
S.F. and E.S. have been supported in part by the D.O.E. 
grant DE-FG03-91ER40662,
Task C and S.F. by the European Community's Human Potential Program 
under contract HPRN-CT-2000-00131
``Quantum Space-Time".  E.I., O.L. and B.Z. are grateful to the DFG grant
No 436 RUS 113/669-02 and a grant of the Heisenberg-Landau program.
The work of O.L. receives support from the DFG grant LE 838/7-2
in the priority programm ``String Theory'' (SPP 1096). 
E.I. and B.Z. also acknowledge support 
from the RFBR grant No 03-02-17440 and the NATO grant PST.GLG.980302.
They thank the Institute of Theoretical Physics of the University of Hannover 
for the kind hospitality extended to them on different stages of this 
work.

\renewcommand\theequation{A.\arabic{equation}} \setcounter{equation}0
\section*{ Appendix: Euclidean harmonic superspace}
We use the conventions
\bea
&&\varepsilon_{12}=-\varepsilon^{12}=
\varepsilon_{\dot{1}\dot{2}}=-\varepsilon^{\dot{1}\dot{2}}=1 \,,
\q \ve^\ab \ve_{\b\g}=\delta^\a_\g \,,
\nn
&&
(\sigma_m)_\ada=(\ii\overrightarrow\sigma, {\bf 1})_\ada~,\q 
(\bar\sigma_m)^{\da\a}=
\varepsilon^\ab\varepsilon^{\da\db}(\sigma_m)_\bdb\,,\\
&&\si_m\bs_n+\si_n\bs_m=2\delta_{mn}\,,\q \si_{mn}
=\sfrac{\ii}{2}(\si_m\bs_n-\si_n\bs_m)\,,\nn
&&\Tr \si_n\bs_m=2\delta_{mn}\,,\q\Tr(\si_n\bs_m\si_p\bs_r)
=2\delta_{nm}\delta_{pr}-
2\delta_{np}\delta_{mr}+2\delta_{nr}\delta_{pm}-2\ve_{nmpr} \nonumber
\eea
for the basic quantities in Euclidean 4D space.
It is convenient to describe the Q-deformations in the chiral coordinates
$Z_\sL=(x^m_\sL,\tka,\btka)$ of Euclidean $N{=}(1,1)$ superspace. 
The spinor
derivatives in these  coordinates are
\be
D^k_\alpha=\partial_\alpha^k+2\ii\bar\theta^{\dot\alpha k}
\partial_{\alpha\dot\alpha}~,\q
\bar D_{\dot\alpha k}=-\bar\partial_{\dot\alpha k}
\ee
while the supersymmetry generators can be chosen as follows,
\be
Q^k_\alpha=\partial^k_\alpha~,\q\bar Q_{\dot\alpha k}=
-\bar\partial_{\dot\alpha k}+2\ii
\theta^\alpha_k\partial_{\alpha\dot\alpha}\lb{QLgen}
\ee
where $\partial_{\alpha\dot\alpha}=(\sigma^m)_{\alpha\dot\alpha}\partial_m$.
The partial harmonic derivatives
\bea
&&\partial^{++}u^+_i=0~,\q \partial^{++}u^-_i=u^+_i,\q \pa^0u^\pm_i
=\pm u^\pm_i\,,\nn
&&\partial^{--}u^+_i=u^+_i~,\q\partial^{--}u^-_i=0,\q [\pa^{++},\pa^{--}]
=\pa^0
\eea
are covariant with respect to supersymmetry
transformations in these coordinates,
\be
[\partial^{\pm\pm},Q^k_\alpha]=0~,\qquad
[\partial^{\pm\pm},\bar Q_{\dot\alpha k}]=0~.
\ee
Note that the chiral $N{=}(1,1)$ superfield can be chosen as real (see, e.g.,
\cite{ILZ}).

The basic concepts of the $N{=}2, D{=}4$ harmonic superspace
are collected in the book \cite{GIOS}. The spinor SU(2)/U(1) harmonics
$u^\pm_i$
can be used to construct analytic coordinates
$(x_\A^m,\theta^{\pm\alpha},\bar\theta^{\pm\da},u^\pm_i)$
in the Euclidean version of $N{=}(1,1)$ harmonic superspace \cite{ILZ,FS}:
\bea
&& x^m_\A=
x^m_\sL-2\ii(\sigma^m)_\ada\theta^{\alpha k}\btja u^{-}_ku^+_j ,\nn
&&\theta^{\alpha\pm}= \theta^{\alpha k}u^\pm_k\ ,\ \q
\bar\theta^{\pm\da} = \bar\theta^{\da k}u^\pm_k\ .
\lb{ancoor}
\eea
The supersymmetry-preserving spinor and harmonic derivatives have
the following form in these coordinates:
\bea
&&\Dpa =\partial_{-\alpha} \ ,\qq
\Dma =-\partial_{+\alpha} + 2\ii\btma\partial_{\alpha\dot\alpha}\ ,\nn [6pt]
&&\bDpa =\bar\partial_{-\da} \ ,\qq
\bDma =-\bar\partial_{+\da} - 2\ii\tma\partial_{\alpha\dot\alpha}\ ,
\lb{Aspder}\\[6pt]
&& \Dp_\A=\dpp -2 \ii\tpa \btpa\partial_{\alpha\dot\alpha}
+ \tpa \partial_{-\alpha} + \btpa\bar\partial_{-\da}
\lb{Aharder}
\eea
where $\partial_{\pm\alpha}\equiv\partial/\partial\theta^{\pm\alpha}$,
$\bar\partial_{\pm\da}\equiv\partial/\partial\bar\theta^{\pm\da}$.
In the analytic coordinates, the $N{=}(1,0)$ supersymmetry generators are
\bea
&&Q^k_\alpha=u^+_k Q^-_\alpha -u^-_k Q^+_\alpha~,\nn
&&Q^+_\alpha=\partial_{-\alpha}-2\ii\bar\theta^{+\da}\pada~,\q
Q^-_\alpha=-\partial_{+\alpha}~.\lb{QAgen}
\eea
The Grassmann-analytic superfields are treated as unconstrained functions
of the coordinates
\be
\zeta=(x^m_\A, \tpa, \btpa)\qquad\textrm{and}\qquad u^\pm_i\,.\lb{zeta}
\ee

We shall also consider the combined chiral-analytic (C) coordinates
$Z_\C=(z_\C,\bar\theta^{\pm\dot\alpha})$ where the chiral part is
\be
z_\C=(x_\sL^m,
\theta^{\pm\alpha})\,. \lb{Ccoor}
\ee
The basic derivatives
have the following form in these coordinates:
\bea
&&\Dpa=\partial_{-\alpha}+ 2\ii\btpa\pada~,\q
\Dma=-\partial_{+\alpha} +2\ii\btma\pada~,\nn
&&\bDpa=\bar\partial_{-\da} ~,\q\bDma=-\bar\partial_{+\da}~,\\
&&\Dp_\C=\partial^{++} +\tpa \partial_{-\alpha} + \btpa\bar\partial_{-\da}~,\nn
&&\Dm_\C=\partial^{--} +\tma \partial_{+\alpha} + \btma\bar\partial_{+\da}~,
\eea
and the $N{=}(1,0)$ supersymmetry generators are
\be
Q^+_\alpha=\partial_{-\alpha}\,,\q Q^-_\alpha=-\partial_{+\alpha}\,.\lb{QCgen}
\ee

The analytic superfields have a decomposition in the C-coordinates:
\be
\L(\zeta,u)=\L(\zeta_\C,u)-2\ii(\tm\si_m\btp)\pa_m\L(\zeta_\C,u)
-(\tm)^2(\btp)^2\Box\L(\zeta_\C,u)\lb{leftanal}
\ee
where $\zeta_\C=(x^m_\sL,\tpa,\btpa)$ and the components of $\L(\zeta_\C,u)$
depend on the left vector coordinates $x^m_\sL$.

In the analytic and C-coordinates, our conventions are
\bea
&&(\tp)^2=\tpa\tp_\a\,,\q (\btp)^2=\btp_\da\btpa\,,\q \tm\si_{mn}\tp
=\theta^{-\alpha}(\si_{mn})_\alpha^\beta \theta^+_\beta
=-\tp\si_{mn}\tm\,,\nn
&&\int \!\diff^4\th\ (\tp)^2(\tm)^2=1,\q \int \!\diff u =1.
\eea
The U(1)-charge operator reads
\be
[\Dp,\Dm]=D^0=\pa^0+\tpa\pa_{+\a}+\btpa\bar\pa_{+\da}
-\tma\pa_{-\a}-\btma\bar\pa_{-\da}\,.
\ee

\vspace{0.2cm}
\noindent{\bf Relations in deformed U(1) gauge theory}

\noindent
Let us rewrite the operator relation \p{ansatz} as
\bea
&&L^{-2}\cA=W_0(x_\sL, \tp,L^{-1}\tm,u)=R_\th 
W_0(x_\sL, \tp,\tm, u)\,, \quad L \equiv 1 +4I\,\bph\,, \nn
&&R_\th=1+(L^{-1}{-}1)(\tm \pa_-)-\sfrac14(L^{-1}{-}1)^2(\tm)^2(\pa_-)^2\,,
\lb{Roper}
\eea
via the rescaling operator for $\tma$,
\be
R_\th \tma=L^{-1}\tma,\q R_\th (\tm)^2=L^{-2}(\tm)^2,\q R_\th\tpa=\tpa.
\ee
Now one can use this operator in the formula
\bea
\cA^2&=& L^4W^2_0(\tp, L^{-1}\tm)=L^4R_\th W_0^2(\tp,\tm)\nn
&=&L^2W^2_0+\pa_{-\a}L^4[1-L^{-1}-\sfrac12(L^{-1}{-}1)^2]\tma W^2_0\nn
&&+\,\sfrac14\pa_{-\a}L^4(L^{-1}-1)^2(\tm)^2\pa^\a_- W^2_0\,, \nn
(\tm \pa_-)W^2_0&=&-\pa_{-\a}(\tma W^2_0)+2W^2_0\,, \nn
(\tm)^2(\pa_-)^2W_0^2&=&\pa_{-\a}[2\tma-(\tm)^2\pa^\a_-]W^2_0
-4W^2_0\,,
\eea
in order to prove \p{U1rel}.

\vspace{0.2cm}
\noindent{\bf Relations in U(n) gauge theory}

\noindent
The undeformed $N{=}(1,0)$ supersymmetry transformations of the elementary 
and composite
component fields of the chiral U($n$) superfield $A$ in \p{Unundef}, \p{Acomp}
have the following form:
\bea
&&\d_\eps\vp=2\eps^{\a k}\j_{\a k},\q\d_\eps\j^k_\a=-\eps_{\a l}d^{kl}
+\sha\eps^k_\b f^\b_\a +\sha\eps^k_\a[\bph,\vp]\,,\nn
&&\d_\eps a_m=\eps^k\si_m\bj_k,\q\d_\eps f_{mn}=\eps^k\si_n\na_m\bj_k-
\eps^k\si_m\na_n\bj_k\,,\nn
&&\d_\eps f_\a^\b=-2\ii
(\eps_{\a k}\xi^{\b k}+\eps^\b_k\xi^k_\a),\q f^\b_\a
=(\si_{mn})^\b_\a f_{mn}\,,\nn
&&\d_\eps d^{kl}=\ii\eps^{\a k}\xi_\a^l+\ii\eps^{\a l}\xi_\a^k
-\eps^{\a k}[\bph,\j^l_\a]-\eps^{\a l}[\bph,\j^k_\a]\,,\nn
&&\d_\eps\bj^{\da k}=\ii\eps_\a^k\na^\ada\bph,\q
\d_\eps \xi^k_\a=\ii\eps^k_\a(p-\sha[\bph,[\bph,\vp]])
-\sfrac{\ii}{2}\eps^k_\b f^\b_\a\,,\nn
&&\d_\eps p=\eps^\a_k\bigl[\bph\,,\,\ii\xi^k_\a-[\bph,\j^k_\a]\bigr]\,.
\lb{10un}
\eea
All fields have the standard gauge transformations, e.g. 
$\d_a a_m =\pa_ma+\ii[a_m,a]$.


\end{document}